\documentclass[12pt,a4paper]{article}
\usepackage{graphicx}
\usepackage[T1]{fontenc}
\usepackage[utf8]{inputenc}
\usepackage{textcomp}
\usepackage[sc,osf]{mathpazo}
\usepackage{a4wide}  
\usepackage{latexsym,amsthm,amsfonts,amsmath,mathrsfs,amssymb}
\usepackage{dsfont}
\usepackage{accents}
\usepackage[nosort]{cite}
\usepackage{booktabs} 
\usepackage[unicode,implicit]{hyperref}
\hypersetup{%
  pdftitle    = {On the thermodynamics of the black holes of the Cano-Ruiperez
    4'dimensional string effective action}
  pdfkeywords = {gravity, string theory, effective action, black hole, thermodynamics},
  pdfauthor   = {Tomas Ortin and Matteo Zatti},
  plainpages  = true,
  colorlinks  = true,
  citecolor   = blue,
  urlcolor    = red,
  linkcolor   = black
}
\newcommand{\hepth}[1]{{\tt
\href{http://www.arXiv.org/abs/hep-th/#1}{hep-th/#1}}}
\newcommand{\grqc}[1]{{\tt
\href{http://www.arXiv.org/abs/gr-qc/#1}{gr-qc/#1}}}

\newcommand{\arxiv}[1]{{\tt arXiv:\href{http://www.arXiv.org/abs/#1}{#1}}}
\newcommand{\doi}[1]{{\tt DOI:\href{http://dx.doi.org/#1}{#1}}}

\allowdisplaybreaks

\makeatletter
\@addtoreset{equation}{section}
\makeatother

\begin{document}

\begin{flushright}
\small
IFT-UAM/CSIC-24-160\\
MPP-2024-221\\
November 15\textsuperscript{th}, 2024\\
\normalsize
\end{flushright}

\vspace{1cm}

\begin{center}

  {\Large {\bf On the thermodynamics of the black holes of the\\[.5cm]
      Cano-Ruip\'erez 4-dimensional string effective action}}
 
\vspace{1cm}

\renewcommand{\thefootnote}{\alph{footnote}}

{\sl\large Tom\'{a}s Ort\'{\i}n}$^{1,}$\footnote{Email:
  {\tt tomas.ortin[at]csic.es}}
{\sl\large and Matteo Zatti}$^{1,2,}$\footnote{Email:
  {\tt zatti[at]mpp.mpg.de}}

\setcounter{footnote}{0}
\renewcommand{\thefootnote}{\arabic{footnote}}
\vspace{1cm}

${}^{1}${\it Instituto de F\'{\i}sica Te\'orica UAM/CSIC\\
  C/ Nicol\'as Cabrera, 13--15,  C.U.~Cantoblanco, E-28049 Madrid, Spain}\\

\vspace{0.5cm}

${}^{2}${\it Max Planck Institut f\"ur Physik \\
	Boltzmannstrasse 8, 85748 Garching, Germany}\\
\vspace{1cm}

{\bf Abstract}
\end{center}
\begin{quotation}
  {\small The Cano--Ruip\'erez 4-dimensional string effective action is the
    simplest consistent truncation of the first-order in $\alpha'$ heterotic
    string effective action compactified on T$^{6}$. This theory contains, on
    top of the metric, a dilaton and an axion that couple to the Gauss-Bonnet
    term and to the Pontrjagin density which suggests a very strong relation
    between the physical and geometrical properties of its
    solutions. In this paper we study the thermodynamics of the string
    black-hole solutions of this theory using Wald's formalism. We construct
    the on-shell closed generalized Komar, dilaton and axion 2-form charges,
    and we used them to find the Smarr formula (that we test on the
    Schwarzschild and Kerr solutions) and to prove no-(primary)-hair theorems
    for the scalar. We find that $\alpha'$ plays the role of a thermodynamical
    variable with an associated chemical potential. We also notice the
    occurrence of non-standard gravitational charges in the Wald entropy and
    scalar charges.
  }
\end{quotation}

\newpage
\pagestyle{plain}



\section{Introduction}

In Ref.~\cite{Cano:2021rey} Cano and Ruip\'erez derived a simple, but very
rich, form of the consistently truncated, first-order in $\alpha'$, heterotic
string effective action compactified on T$^{6}$. Since all the vector fields
and most of the scalars were truncated, this theory is useful to analyze the
$\alpha'$ corrections to uncharged solutions. While the corrections to the
Kerr black-hole solution were already computed in that reference, some
important aspects of the thermodynamics of the $\alpha'$-corrected solutions
are yet to be explored. This is the main goal of this paper, but, in our way
to achieve it, we are going to have to meet a number of problems for which we
have found sound but intriguing solutions.

In order to better explain these problems, let us introduce the
Cano--Ruip\'erez string effective action first.

\subsection{The Cano-Ruip\'erez (CR) string effective action}
\label{sec-thetheory}

Compactifying on T$^{6}$ the first order in $\alpha'$, 10-dimensional,
Bergshoeff-de Roo heterotic string effective action \cite{Bergshoeff:1989de},
consistently truncating the gauge fields (and some of the scalar fields) and
performing several field redefinitions, Cano and Ruip\'erez found in
Ref.~\cite{Cano:2021rey} the following 4-dimensional effective action for the
metric (which we are going to describe through the Vierbein
$e^{a}=e^{a}{}_{\mu}dx^{\mu}$), dilaton $\phi$ and axion\footnote{The
  4-dimensional axion field is the dual of the Kalb--Ramond 2-form.} $\chi$
fields\footnote{In this paper we are going to use differential forms and the
  conventions of Refs.~\cite{Ortin:2015hya,Gomez-Fayren:2023qly}. Some of them
  are described in Appendix~\ref{sec-conventions}.}:
 
\begin{equation}
  \label{eq:CRaction}
\begin{aligned}
  S_{\rm CR}[e^{a},\phi,\chi]
  & =
    \frac{1}{16\pi G_{N}} \int \left\{ -\star(e^{a}\wedge e^{b}) \wedge R_{ab}
    +\tfrac{1}{2}\left(d\phi \wedge \star d\phi
    +e^{2\phi} d\chi \wedge \star d\chi\right)
    \right.
  \\
    & \\
  & \hspace{.5cm}
    \left.
    -\frac{\alpha'}{4}\left(e^{-\phi} \tilde{R}^{ab}\wedge R_{ab}
    -\chi R^{a}{}_{b}\wedge R^{b}{}_{a}\right) \right\}\,,
\end{aligned}
\end{equation}

\noindent
where

\begin{equation}
  \tilde{R}^{ab}
  \equiv
  \tfrac{1}{2}\varepsilon^{abcd}R_{cd}\,.
\end{equation}

We will refer to this action as the Cano--Ruip\'erez (CR) action.

The Lagrangian can be split into a zeroth- and  a first-order term

\begin{subequations}
  \begin{align}
    \mathbf{L}
    & =
      \mathbf{L}^{(0)}+\mathbf{L}^{(1)}\,,
    \\
    & \nonumber \\
    \label{eq:L0}
    \mathbf{L}^{(0)}
    & =
      -\star(e^{a}\wedge e^{b}) \wedge R_{ab}
      +\tfrac{1}{2}\left(d\phi \wedge \star d\phi
      +e^{2\phi} d\chi \wedge \star d\chi\right)\,,
    \\
    & \nonumber \\
        \label{eq:L1}
    \mathbf{L}^{(1)}
    & =    
      -\frac{\alpha'}{4}\left(e^{-\phi} \tilde{R}^{ab}\wedge R_{ab}
      -\chi R^{a}{}_{b}\wedge R^{b}{}_{a}\right)\,.    
  \end{align}
\end{subequations}

$\mathbf{L}^{(0)}$ is the action of a SL$(2,\mathbb{R})/$SO$(2)$
$\sigma$-model coupled to gravity. 

In $\mathbf{L}^{(1)}$ we find two topological invariants multiplied by the
scalars of the theory:

\begin{enumerate}
\item The Gauss--Bonnet (GB) 4-form

\begin{equation}
  2\tilde{R}^{ab}\wedge R_{ab}
  =
  R^{ab}\wedge R^{cd}\varepsilon_{abcd}\,,  
\end{equation}

\noindent
multiplying the exponential of the dilaton.

\item The Pontrjagin (P) 4-form

  \begin{equation}
R^{a}{}_{b}\wedge R^{b}{}_{a}\,,  
\end{equation}

\noindent
multiplying the axion.

\end{enumerate}

If we set to zero the axion, $\chi=0$, the CR action becomes that of the very
popular dilaton-Gauss-Bonnet theory, which is commonly understood as a
string-inspired theory. However, as stressed in Ref.~\cite{Cano:2021rey},
setting to zero the axion is not a consistent truncation of the string
effective action and, in general, only the solutions of the above theory can
be embedded in string theory. If we, somehow, could get rid of the dilaton and
the GB term we would get the ``dynamical Chern--Simons modified gravity'' of
Ref.~\cite{Sopuerta:2009iy} which is also ``string-inspired'' but is certainly
not a consistent truncation of the string effective action either.  

The P 4-form is a total derivative in any dimension:

\begin{subequations}
  \begin{align}
    R^{a}{}_{b}\wedge R^{b}{}_{a}
    & =
      d\omega_{(3)}^{L}\,,
    \\
    & \nonumber \\
    \omega_{(3)}^{L}
    & =
      d\omega^{a}{}_{b}\wedge \omega^{b}{}_{a}
      -\tfrac{2}{3}\omega^{a}{}_{b}\wedge \omega^{b}{}_{c}
      \wedge \omega^{c}{}_{a}
    \nonumber \\
    & \nonumber \\
     &  =
      R^{a}{}_{b}\wedge \omega^{b}{}_{a}
      +\tfrac{1}{3}\omega^{a}{}_{b}\wedge \omega^{b}{}_{c}
      \wedge \omega^{c}{}_{a}\,,
  \end{align}
\end{subequations}

\noindent
where $\omega_{(3)}^{L}$ is the Lorentz Chern--Simons 3-form.

The GB 4-form can only be properly defined in 4 dimensions, but there are
obvious generalizations in higher dimensions. Its variation is a total
derivative

\begin{equation}
   \tilde{R}^{ab}\wedge \delta R_{ab}
  =
  d\left(2 \tilde{R}^{ab}\wedge \delta \omega_{ab}\right)\,,
\end{equation}

\noindent
and, therefore, its normalized integral on 4-dimensional closed manifolds is a
topological invariant known as the Euler--Poincar\'e characteristic, according
to the Chern--Gauss--Bonnet theorem. If the manifold is not closed (as it
usually is the case in the black-hole physics context, a boundary term needs
to be added to the GB 4-form in order to obtain a topological invariant that
gives the Euler--Poincar\'e characteristic. See Ref.~\cite{Gibbons:1994ff} and
references therein.

\begin{equation}
  \chi(\mathcal{M})
  \equiv
  \tfrac{1}{16\pi^{2}}\int_{\mathcal{M}}\tilde{R}^{ab}\wedge R_{ab}\,.  
\end{equation}

A more standard expression of the GB 4-form is

\begin{equation}
2  R^{ab}\wedge \tilde{R}_{ab}
  =
  -\sqrt{|g|}\left[R_{abcd}R^{abcd}-4 R_{\mathrm{ic}\,ab}R_{\mathrm{ic}}{}^{ab}+R_{\mathrm{ic}}^{2}\right]\,.
\end{equation}

Both topological terms only depend on the Vierbein through the Levi--Civita
spin connection $\omega^{ab}=\omega^{ab}(e)$. They do not contain the Vierbein
explicitly.

The CR action bears some resemblance to that of the so-called axion-dilaton
gravity, in which the two invariants of the Lorentz curvature are replaced by
two invariants of Maxwell fields

\begin{equation}
  F\wedge \star F\,,
  \,\,\,\,\,
  \text{and}
  \,\,\,\,\,
  F\wedge F\,,
\end{equation}

\noindent
respectively. In both cases, the kinetic terms of the scalars correspond to a
SL$(2,\mathbb{R})/$SO$(2)$ $\sigma$-model, but only the equations o motion of
the axion-dilaton theory are SL$(2,\mathbb{R})$ invariant.\footnote{The action
  of the axion-dilaton theory is only invariant under constant shifts of the
  axion (up to total derivatives) and of the dilaton. The electric-magnetic
  duality part of SL$(2,\mathbb{R})$ is only a symmetry of the equations of
  motion.} The CR action is invariant under constant shifts of the axion up to
total derivatives, but not under shifts of the dilaton, unless we rescale
$\alpha'$. A rescaling of $\alpha'$ is not a field transformation and would
take us to another theory with the same form but with a different value of
that constant. Thus, at least in this setting one can say that $\alpha'$
corrections break the shift symmetry of the dilaton.

Our main goal is to study the thermodynamics of the black holes of the CR
theory. In particular, we want to derive the Smarr formula of the black holes
of this theory using Wald's formalism
\cite{Lee:1990nz,Wald:1993nt,Iyer:1994ys}, which, following
Refs.~\cite{Kastor:2008xb,Kastor:2010gq}, requires the calculation of the
generalized Komar charge 2-form along the lines of
Refs.~\cite{Liberati:2015xcp,Ortin:2021ade,Mitsios:2021zrn}.

As argued in Ref.~\cite{Ortin:2021ade} (see also Ref.~\cite{Hajian:2021hje}),
in the CR theory the Komar charge 2-form and, hence, the Smarr formula, should
contain a term in which the dimensionful constant $\alpha'$ plays the role of
a thermodynamical variable. In order to obtain this term, it is necessary to
promote the dimensionful constant to a scalar field, adding a
Lagrange-multiplier term that forces it to be constant on-shell
\cite{Meessen:2022hcg}. The Lagrange multiplier is a 3-form field and the new,
on-shell-equivalent theory is invariant under shifts of the 3-form by the
exterior derivative of arbitrary 2-forms. When this gauge symmetry is
correctly accounted for in Wald's formalism as in
Refs.~\cite{Elgood:2020svt,Elgood:2020mdx,Elgood:2020nls}, there is a new term
in the Noether--Wald charge and generalized Komar 2-forms which lead to the
term we were looking for in the Smarr formula.

This additional term has been found in the Smarr formulas of
$\alpha'$-corrected heterotic string black-hole solutions \cite{Cano:2022tmn, Zatti:2023oiq}
and it is expected to be present in the Smarr formula of the CR theory as
well. The need for this term is easy to see if we consider the
$\alpha'$-corrected Schwarzschild black hole in the CR theory: at first order,
only the scalars get corrections. Since the metric is not corrected, neither
the mass, nor the temperature will get corrections $M=M_{0},T=T_{0}$. The
entropy will get corrections, though, because now it is not just given by the
Bekenstein--Hawking term and it will contain new terms given by the Iyer--Wald
prescription \cite{Iyer:1994ys}. Thus, $S=S_{0}+\alpha' S_{1}$. The standard
Smarr formula \cite{Smarr:1972kt} derived from the Komar charge
\cite{Komar:1958wp} has the form ($J=0$)

\begin{equation}
M = 2TS\,,  
\end{equation}

\noindent
and it is identically satisfied at zeroth order:

\begin{equation}
M_{0} = 2T_{0}S_{0}\,.  
\end{equation}

\noindent
As we have discussed, at first order only $S$ is corrected and, therefore, it
cannot be satisfied. Then, there must be an additional term
$\alpha'\Phi_{\alpha'}$ in the right-hand side to compensate the term
$\alpha' T_{0}S_{1}$, $\Phi_{\alpha'}= - T_{0}S_{1}$ being the chemical
potential associated to the thermodynamical variable (or charge) $\alpha'$.

All this leads us to consider the on-shell-equivalent theory in which
$\alpha'$ is treated as a scalar field, since it is the only way in which the
additional term we know must be present in the Smarr formula can be
found.

This on-shell-equivalent theory that we will simply call CR2 is, however, more
general, and more interesting than the original one. First of all, since
$\alpha'$ is now a field $\alpha'(x)$ we can rescale it and, as we are going
to see, the invariance of the theory under dilaton shifts is restore. Somewhat
miraculously, the invariance under axion shifts which is apparently broken by
the spacetime-dependent $\alpha'$ is, nevertheless, preserved. 

The rest of the paper is organized as follows: in
Section~\ref{sec-thetheory2} we introduce the CR2 action, computing its
equations of motion ad presymplectic potential. In Section~\ref{sec-global} we
study the global symmetries of the CR2 action and the associated conserved
charges. In Section~\ref{sec-local} we study the local symmetries of the CR2
action (local Lorentz transformations and diffeomorphisms) and the associated
Noether charges. In Section~\ref{sec-scalarcharges} we define on-shell-closed
2-form charges associated to the global symmetries that act on the scalars in
the CR2 action and we study the consequences of the existence of these
charges.  In Section~\ref{sec-Komar} we derive the generalized Komar charge of
the CR2 action and the Smarr formula for the black hole of the theory,
comparing with the results for the $\alpha'$-corrected Schwarzschild and Kerr
black holes. Section~\ref{sec-discussion} contains our conclusions and some
questions we would like to answer in future work.

\section{The Cano--Ruip\'erez modified action (CR2)}
\label{sec-thetheory2}

It is convenient to replace $\alpha'$ by the string length $\ell_{s}$
($\alpha'=\ell_{s}^{2}$). Now $\ell_{s}=\ell_{s}(x)$ will be treated as a
scalar field and we introduce a Lagrange-multiplier 3-form field $C$
enforcing the constraint $d\ell_{s}=0$.  The action of the CR2 theory is

\begin{equation}
  \label{eq:CR2action}
\begin{aligned}
  S_{\rm CR2}[e^{a},\phi,\chi,\ell_{s},C]
  & =
    \frac{1}{16\pi G_{N}} \int \left\{ -\star(e^{a}\wedge e^{b}) \wedge R_{ab}
    +\tfrac{1}{2}\left(d\phi \wedge \star d\phi
    +e^{2\phi} d\chi \wedge \star d\chi\right)
    \right.
  \\
    & \\
  & \hspace{.5cm}
    \left.
    -\frac{\ell_{s}^{2}}{4}\left(e^{-\phi} R^{ab}\wedge \tilde{R}_{ab}
    -\chi R^{a}{}_{b}\wedge R^{b}{}_{a}\right)
    +\ell_{s}dC\right\}
  \\
  & \\
  & \equiv
    \int \mathbf{L}\,.
\end{aligned}
\end{equation}

By construction, the equation of $C$ is $d\ell_{s}=0$. If we substitute the
solution $\ell_{s}=$ constant in the equations of motion of the Vielbein and
of the scalar fields, they become those of the original CR theory. Notice,
though, that there are other solutions in which $\ell_{s}$ is only piecewise
constant and it takes different constant values in different regions of
spacetime, separated by domain walls. This kind of solutions has never been
considered before and we intend to study them in future work along the lines
of Ref.~\cite{Bergshoeff:2001pv}. For the moment, we would simply like to
stress that this theory is slightly more general than the original one.

In the rest of this section we are going to derive its equations of motion.

\subsection{Equations of motion}
\label{sec-eom}

The CR and CR2 actions contain terms of higher order in derivatives and we
have found it convenient to proceed in two steps. First, we vary the action
with respect to the explicit occurrence of all the fields and with respect to
the occurrences of the fields through their field strengths:

\begin{equation}
  \label{eq:variationSCR2-1}
  \begin{aligned}
  \delta S_{\rm CR2}
  & =
  \int
  \left\{\mathbf{E}^{(0)}{}_{a} \wedge \delta e^{a}
    +\tilde{\mathbf{E}}^{(1)}{}_{ab} \wedge \delta R^{ab}
    +\mathbf{E}^{(0)}{}_{\phi} \, \delta \phi
    +\tilde{\mathbf{E}}^{(1)}{}_{\phi} \, \delta d\phi
    +\mathbf{E}^{(0)}{}_{\chi} \, \delta \chi
    +\tilde{\mathbf{E}}^{(1)}{}_{\chi} \, \delta d\chi
  \right.
  \\
  & \\
  & \hspace{.5cm}
  \left.
    +\mathbf{E}_{\rm \ell_{s}}\delta \ell_{s}
    +\tilde{\mathbf{E}}^{(1)}{}_{\rm C}\delta dC
    \right\}\,.
  \end{aligned}
\end{equation}

The terms in boldface are the building blocks of the equations of motion and
of the presymplectic potential and are given in
Appendix~\ref{app-buildingblocks}.

Using $\delta d = d\delta$ and the Palatini
identity $\delta R^{ab}=\mathcal{D}\delta \omega^{ab}$ and integrating by
parts, we get

\begin{equation}
  \label{eq:variationSCR2-2}
  \begin{aligned}
  \delta S_{\rm CR2}
  & =
  \int
  \left\{\mathbf{E}^{(0)}{}_{a} \wedge \delta e^{a}
    +\mathbf{E}^{(1)}{}_{ab} \wedge \delta \omega^{ab}
    +\mathbf{E}_{\phi} \, \delta \phi
    +\mathbf{E}_{\chi} \, \delta \chi
    +\mathbf{E}_{\rm \ell_{s}}\delta \ell_{s}
        +\mathbf{E}_{\rm C}\wedge \delta C
  \right.
  \\
  & \\
  & \hspace{.5cm}
  \left.
    +d\left[ \tilde{\mathbf{E}}^{(1)}{}_{ab} \wedge \delta \omega^{ab}
    -\tilde{\mathbf{E}}^{(1)}{}_{\phi} \, \delta \phi
    -\tilde{\mathbf{E}}^{(1)}{}_{\chi} \, \delta \chi
    +\tilde{\mathbf{E}}^{(1)}{}_{\rm C}\delta C
    \right]
    \right\}\,,
  \end{aligned}
\end{equation}

\noindent
where we have defined

\begin{subequations}
  \begin{align}
    \label{eq:E1abdef}
  \mathbf{E}^{(1)}{}_{ab}
  & \equiv
    -\mathcal{D}\tilde{\mathbf{E}}^{(1)}{}_{ab}\,,
    \\
    & \nonumber \\
    \mathbf{E}_{\phi}
    & \equiv
      \mathbf{E}^{(0)}{}_{\phi} 
      +d\tilde{\mathbf{E}}^{(1)}{}_{\phi}\,,
    \\
    & \nonumber \\
    \mathbf{E}_{\chi}
    & \equiv
      \mathbf{E}^{(0)}{}_{\chi} 
      +d\tilde{\mathbf{E}}^{(1)}{}_{\chi}\,,
    \\
    & \nonumber \\
      \mathbf{E}_{\rm C}
    & \equiv
      -d\tilde{\mathbf{E}}^{(1)}{}_{\rm C}\,,
  \end{align}
\end{subequations}

The explicit form of $\mathbf{E}^{(1)}{}_{ab}$ can be found in
Eq.~(\ref{eq:E1abexplicit}).

The variation of the Levi--Civita action in terms of the variation of the
Vielbein is

\begin{equation}
  \label{eq:deltaomegadeltae}
  \delta \omega_{ab}
  =
  \tfrac{1}{2} \left\{\imath_{c} \imath_{a} \mathcal{D} \delta e_{b}
    -\imath_{a} \imath_{b} \mathcal{D} \delta e_{c}
    +\imath_{b} \imath_{c} \mathcal{D} \delta e_{a} \right\} e^{c} \,,
\end{equation}

\noindent
and

\begin{equation}
  \begin{aligned}
    \int \mathbf{E}_{ab}^{(1)} \wedge \delta \omega^{ab}
    & =
      \int 
      \left\{
      -\mathcal{D}\Delta_{c}
      +d\left(\Delta_{c}\wedge \delta e^{c}\right)\right\}\,,
  \end{aligned}
\end{equation}

\noindent
where

\begin{equation}
    \label{eq:Deltac}
    \Delta_{c}
    \equiv
    2\imath_{a}\mathbf{E}^{(1)\, a}{}_{c}
    +\tfrac{1}{2} \imath_{b}\imath_{a}\mathbf{E}^{(1)\, ab}
    \wedge e^{d}g_{dc}\,,  
\end{equation}

\noindent
and, substituting this result in Eq.~(\ref{eq:variationSCR2-2}), we finally
get

\begin{equation}
  \label{eq:variationSCR2-3}
  \delta S_{\rm CR2}
   =
  \int
  \left\{\mathbf{E}_{a} \wedge \delta e^{a}
    +\mathbf{E}_{\phi} \, \delta \phi
    +\mathbf{E}_{\chi} \, \delta \chi
    +\mathbf{E}_{\rm \ell_{s}}\delta \ell_{s}
        +\mathbf{E}_{\rm C}\wedge \delta C
    +d\mathbf{\Theta}(\varphi,\delta\varphi)
    \right\}\,,
\end{equation}

\noindent
where the Einstein equation and symplectic prepotential are respectively given
by

\begin{subequations}
  \begin{align}
    \label{eq:Ea}
    \mathbf{E}_{a}
    & =
      \mathbf{E}^{(0)}{}_{a}
      -\mathcal{D}\Delta_{a}\,,
      \\
    & \nonumber \\
    \label{eq:presymplecticpotential}
    \mathbf{\Theta}(\varphi,\delta\varphi)
    & =
      \tilde{\mathbf{E}}^{(1)}{}_{ab} \wedge \delta \omega^{ab}
    -\tilde{\mathbf{E}}^{(1)}{}_{\phi} \, \delta \phi
    -\tilde{\mathbf{E}}^{(1)}{}_{\chi} \, \delta \chi
      +\tilde{\mathbf{E}}^{(1)}{}_{\rm C}\delta C
      +\Delta_{a}\wedge \delta e^{a}\,.
  \end{align}
\end{subequations}

The explicit form of the presymplectic potential is in
Eq.~(\ref{eq:presymplecticpotentialexplicit}) and the explicit form of all the
equations of motion can be found in Appendix~\ref{app-eoms}.

Eq.~(\ref{eq:Deltac}) leads to the identity

\begin{equation}
  \label{eq:identity}
  \mathbf{E}^{(1)\, ab}
  =
  -\Delta^{[a}\wedge e^{b]}\,,
\end{equation}

\noindent
which can be easily derived (in any dimension $d$) from

\begin{equation}
  \tfrac{1}{2}\imath_{b} \imath_{a}
  \left(\mathbf{E}^{(1)\,ab} \wedge e^{c} \wedge e^{d} \right) = 0\,,
\end{equation}

\noindent
and the definition. Observe that this identity is based on the fact that the
expression in parenthesis is a $(d+1)$-form and it is equivalent to the use of
a Schouten identity.

We notice that the definitions of
$\tilde{\mathbf{E}}^{(1)\,ab},\mathbf{E}^{(1)\,ab},\Delta^{a}$ and the
property Eq.~(\ref{eq:identity}) are completely generic and not just a
property of the CR or CR2 theories.

\section{Global symmetries}
\label{sec-global}

The original CR action Eq.~(\ref{eq:CRaction}) is not invariant under constant
shifts of the dilaton, but the modified action CR2 Eq.~(\ref{eq:CR2action}) is
invariant under 

\begin{equation}
  \delta_{m}\phi = m\,,
  \hspace{.5cm}
  \delta_{m}\chi = -m \chi\,,
  \hspace{.5cm}
  \delta_{m}\ell_{s}= \frac{m}{2}\ell_{s}\,,
  \hspace{.5cm}
  \delta_{m}C = -\frac{m}{2}C\,,
\end{equation}

\noindent
where $m$ is an infinitesimal, real parameter.

Observe that the rescaling of $\ell_{s}$ (hence, of $\alpha'$) is a legitimate
transformation of this action because in it $\ell_{s}$ is a field.

The Noether current associated to this global symmetry of the CR2 action is 

\begin{equation}
  J_{m}  \equiv  \star d\phi  -e^{2\phi}\star d\chi -\tfrac{1}{2}\ell_{s}C\,,
  \hspace{1cm}
  dJ_{m}
  \doteq
  0\,.
\end{equation}

At first sight, when $\ell_{s}$ is promoted to a field, the invariance up to
total derivatives of the original CR action under constant shifts of the axion
gets broken. However, it is not difficult to see that the CR2 action is
invariant under the infinitesimal transformations

\begin{equation}
  \delta_{n}\chi = n\ell_{s}^{-2}\,,
  \hspace{1cm}
  \delta_{n}C = -2 n\ell_{s}^{-3} e^{2\phi} \star d\chi\,,
\end{equation}

\noindent
up to a total derivative given by

\begin{equation}
  \delta_{n}S_{\rm CR2}
  =
  n \int d\left[\tfrac{1}{4}\omega_{(3)}^{L}
    -2\ell_{s}^{-2}e^{2\phi}\star d\chi\right]\,.
\end{equation}

The associated Noether current is given by

\begin{equation}
  J_{n}
  =
  \ell_{s}^{-2}e^{2\phi}\star d\chi - \tfrac{1}{4}\omega_{(3)}^{L}\,,
  \hspace{1cm}
  dJ_{n}
  \doteq
  0\,.
\end{equation}

As shown in Ref.~\cite{Ballesteros:2023iqb}, in static black-hole spacetimes,
the charges obtained by integrating these 3-forms over spacelike hypersurfaces
vanish identically. In order to characterize the scalar fields one can define,
instead, 2-form charges which are derived from the above 3-form currents by
exploiting the assumption of invariance under the diffeomorphisms generated by
a Killing vector $k$. We will do this in Section~\ref{sec-scalarcharges}, but
we have to discuss how the different fields transform under local Lorentz
transformations and diffeomorphisms first.

\section{Local symmetries}
\label{sec-local}

\subsection{Local Lorentz transformations}

We start considering the local Lorentz transformations. Only the Vierbein and
the fields derived from it (spin connection and curvature) have non-trivial
local Lorentz transformations. They take the form

\begin{equation}
	\delta_{\sigma} e^{a}
	=
	\sigma^{a}{}_{b} e^{b}\,, \qquad
	\delta_{\sigma} \omega_{ab}
	=
	\mathcal{D} \sigma_{ab} \,, \qquad
	\delta_{\sigma} R^{ab}
	=
	2\sigma^{[a|}{}_{c}R^{c|b]}\,. 
\end{equation} 

\noindent
The CR2 action is exactly invariant under this transformation. We obtain

\begin{equation}
	0 = \delta_{\sigma} S_{\rm CR2}
	=
	\int\left\{ \mathbf{E}^{a} \wedge e^{b}\sigma_{ab} 
	+d \mathbf{\Theta}(\varphi,\delta_{\sigma}\varphi)\right\}\,,
\end{equation}

\noindent
which implies the following (off-shell) Noether identity

\begin{equation}
	\mathbf{E}^{ [a}\wedge e^{b]}  = 0 \,,
\end{equation}

\noindent
and the off-shell-closed Noether current $\mathbf{J}[\sigma]$

\begin{equation}
	\mathbf{J}[\sigma]
		 \equiv
	\mathbf{\Theta}(\varphi,\delta_{\sigma}\varphi) 
		 =
		d{\mathbf{Q}}[\sigma]\,,
\end{equation}

\noindent
where

\begin{equation}
  \label{eq:localLorentzcharge}
	\mathbf{Q}[\sigma]
	\equiv
	-\star (e^{a} \wedge e^{b})\sigma_{ab} 
	-\tfrac{1}{2}\ell_{s}^{2}\left(e^{-\phi} \tilde{R}^{ab}\sigma_{ab}
	+\chi R^{ab}\sigma_{ab}\right)\,,
\end{equation}

\noindent
is the Noether 2-form charge associated to local Lorentz invariance. In order
to perform this computation it is convenient to use the identity
Eq.~(\ref{eq:identity}).

By construction, this charge is closed on-shell for local Lorentz parameters
that leave the connection invariant, \textit{i.e.}~which are covariantly
constant. As we have stressed elsewhere, the invariance of the Vierbein is not
required. Furthermore, as we will see, Wald's black-hole entropy is actually
given by this charge, for a particular Lorentz parameter.

\subsection{Gauge transformations of the auxiliary 3-form field $C$}

The CR2 action is exactly invariant under the transformations

\begin{equation}
  \delta_{\Lambda} C
  =
  d\Lambda\,,
\end{equation}

\noindent
where $\Lambda$ is any 2-form. Then, from the general variation of the CR2
action Eq.~(\ref{eq:variationSCR2-3}), we get

\begin{equation}
  0
  =
  \delta_{\Lambda}S_{\rm CR2}
  =
  \int \left\{\mathbf{E}_{C}\wedge d\Lambda
  +d\mathbf{\Theta}(\varphi,\delta_{\Lambda}\varphi)\right\}\,,
\end{equation}

\noindent
which leads to the very trivial Noether identity $d\mathbf{E}_C=0$ and to the
off-shell closed current

\begin{equation}
  \mathbf{J}[\Lambda]
  =
  \mathbf{\Theta}(\varphi,\delta_{\Lambda}\varphi)-\mathbf{E}_{C}\wedge
  \Lambda
  =
  d\mathbf{Q}[\Lambda]\,,
\end{equation}

\noindent
with

\begin{equation}
  \mathbf{Q}[\Lambda]
  \equiv
  \ell_{s}\Lambda\,.
\end{equation}

On-shell and for Killing (or \textit{reducibility}) gauge parameters $K$
leaving $C$ invariant (\textit{i.e.}~closed, $dK =0$) $\mathbf{Q}[K]$ is a
closed 2-form and the charges obtained by integrating it over closed
2-surfaces satisfy a Gauss law.

\subsection{Diffeomorphisms}

The CR2 action is manifestly invariant under diffeomorphisms.  Thus, under
infinitesimal diffeomorphisms

\begin{equation}
  \delta_{\xi} x^{\mu}
  =
  \xi^{\mu}\,.
\end{equation}

\noindent
represented on the fields by the action of the standard Lie derivative
$\pounds_{\xi}$,

\begin{equation}
  \delta_{\xi}\varphi
  =
  -\pounds_{\xi}\varphi\,,
\end{equation}

\noindent
it is invariant up to a total derivative

\begin{equation}
  \label{eq:totalderivativediffs}
  \delta_{\xi}S
  =
  -\int d\imath_{\xi}\mathbf{L}\,.
\end{equation}

As it has been discussed elsewhere \cite{Jacobson:2015uqa,Elgood:2020svt}, the
transformations that have to be used when fields with Lorentz indices are
involved

\begin{equation}
  \delta_{\xi}\varphi
  =
  -\mathbb{L}_{\xi}\varphi\,,
\end{equation}

\noindent
where $\mathbb{L}_{\xi}$ is the Lie-Lorentz derivative\footnote{See
  Ref.~\cite{Ortin:2002qb} and references therein.} which is essentially a
standard Lie derivative plus a ``compensating'' local Lorentz transformation
with a $\xi$- and field-dependent parameter given by

\begin{equation}
  \sigma_{\xi}^{ab}
  \equiv
  \imath_{\xi}\omega^{ab}-P_{\xi}{}^{ab}\,,
\end{equation}

\noindent
with 

\begin{equation}
  P_{\xi}{}^{ab}
  \equiv
  \mathcal{D}^{[a}\xi^{b]} \,.
\end{equation}

For $p$-form fields with standard gauge transformations one has to take into
account a compensating gauge transformation, which, for the 3-form $C$, has a
2-form gauge parameter $\Lambda_{\xi}$ of the form

\begin{equation}
  \Lambda_{\xi}
  =
  \imath_{\xi}C-P_{\xi}\,,
\end{equation}

\noindent
such that, for Killing vectors $\xi=k$, the 2-form $P_{k}$ satisfies the
momentum map equation

\begin{equation}
  \imath_{k}dC+ dP_{k}
  =
  0\,.  
\end{equation}

Thus, for the fields of the CR2 action, we must consider the transformations

\begin{subequations}
  \label{eq:deltaxi}
  \begin{align}
    \delta_{\xi} e^{a}
    & =
      -\left(\mathcal{D} \xi^{a} + P_{\xi}{}^{a}{}_{b}e^{b}\right) \,,
    \\
    & \nonumber \\
    \delta_{\xi} \omega_{ab}
    & =
      -\left(\imath_{\xi} R_{ab} + \mathcal{D}P_{\xi\, ab}\right) \,,
    \\
    & \nonumber \\
    \delta_{\xi} \phi
    & =
      -\imath_{\xi} d \phi \,,
    \\
    & \nonumber \\
    \delta_{\xi} \chi
    & =
      -\imath_{\xi} d \chi \,,
    \\
    & \nonumber \\
    \delta_{\xi} \ell_{s}
    & =
      -\imath_{\xi} d \ell_{s} \,,
    \\
    & \nonumber \\
    \delta_{\xi}C
    & =
      -\imath_{\xi}dC -dP_{\xi}\,.
  \end{align}
\end{subequations}

\subsection{Noether--Wald Charge}

Since the CR2 action is exactly invariant under all the local symmetries we
have studied, it must be invariant up to a total derivative under the field
transformations Eqs.~(\ref{eq:deltaxi})

\begin{equation}
    \label{eq:deltaxiS-1}
  \delta_{\xi}S_{\rm CR2}
  =
  -\int d\imath_{\xi}\mathbf{L}\,.
\end{equation}

On the other hand, using the general variation of the action
Eq.~(\ref{eq:variationSCR2-3}), integrating by parts, and using the Noether
identities associated to local Lorentz transformations and gauge
transformations of $C$ we arrive to

\begin{equation}
  \label{eq:deltaxiS-2}
  \begin{aligned}
    \delta_{\xi} S_{\rm CR2}
    & =
      -\int\left\{ \mathcal{D} \mathbf{E}_{a}\xi^{a}
      +\mathbf{E}_{\phi} \, \imath_{\xi} d \phi
      +\mathbf{E}_{\chi} \, \imath_{\xi} d \chi
      +\mathbf{E}_{\chi} \, \imath_{\xi} d \ell_{s}
      +\mathbf{E}_{C} \wedge \imath_{\xi} d C 
      -d{\mathbf{\Theta}}'(\varphi,\delta_{\xi}\varphi)
      \right\}\,,
	\end{aligned} 
\end{equation}

\noindent
with

\begin{equation}
  \mathbf{\Theta}'(\varphi,\delta_{\xi}\varphi)
 =
 \mathbf{\Theta}(\varphi,\delta_{\xi}\varphi)
 +\xi^{a} \mathbf{E}_{a} +\mathbf{E}_{C} \wedge P_\xi\,.
\end{equation}

\noindent
The above expression leads to the Noether identity

\begin{equation}
	 \mathcal{D} \mathbf{E}_{a}\xi^{a}
	+\mathbf{E}_{\phi} \, \imath_{\xi} d \phi
	+\mathbf{E}_{\chi} \, \imath_{\xi} d \chi
        +\mathbf{E}_{\chi} \, \imath_{\xi} d \ell_{s}
        +\mathbf{E}_{C} \wedge \imath_{\xi} d C  = 0 \,,
\end{equation}

\noindent
that can be easily checked exploiting the relation

\begin{equation}
  \begin{aligned}
    \mathcal{D} \mathbf{E}^{(1)}{}_{a}\xi^{a}
    & =  
      -\xi^{a}\mathcal{D}\mathcal{D}\Delta_{a}
      =
      \xi^{a} R_{ab}\wedge \Delta^{b}
      =
      \tfrac{1}{2}\xi^{a} R_{cdab} e^{c}\wedge e^{d} \wedge\Delta^{b}
    \\
    & \\
    & =
      -\xi^{a} R_{acdb}e^{c}\wedge e^{d} \wedge\Delta^{b}
      =
      -\imath_{\xi}R_{db}\wedge e^{d} \wedge\Delta^{b}
      =
      \imath_{\xi}R_{ab} \wedge \Delta^{a} \wedge e^{b}
    \\
    & \\
    & =
      -\imath_{\xi}R_{ab} \wedge \mathbf{E}^{(1)\, ab}\,.
	\end{aligned}
\end{equation}

Comparing Eqs.~(\ref{eq:deltaxiS-1}) and (\ref{eq:deltaxiS-2}) 
 we arrive to the off-shell identity

\begin{equation}
d \mathbf{J}[\xi] = 0 \,,
\end{equation}

\noindent
where

\begin{equation}
  \mathbf{J}[\xi]
  \equiv
  \mathbf{\Theta}'(\varphi,\delta_{\xi}\varphi)
  +\imath_{\xi} \mathbf{L}\,.
\end{equation}

$\mathbf{J}[\xi]$ must be locally exact and, indeed, we find that

\begin{subequations}
  \begin{align}
  \mathbf{J}[\xi]
  & =
  d \mathbf{Q}[\xi] \,,
  \\
    & \nonumber \\
    \label{eq:NWcharge}
  \mathbf{Q}[\xi]
  & \equiv
  \star (e^{a} \wedge e^{b})P_{\xi\, ab}
  +\tfrac{1}{2}\ell_{s}^{2}\left(e^{-\phi} \tilde{R}^{ab}
    +\chi R^{ab}\right)P_{\xi\, ab}
  -\ell_{s} P_{\xi} -\Delta_{a} \xi^{a} \,.
  \end{align}
\end{subequations}

$\mathbf{Q}[\xi]$ is the \textit{Noether--Wald} 2-form charge.

Observe that using the Iyer-Wald prescription \cite{Iyer:1994ys} we would
have obtained ($\xi=k$ and $P_{k}{}^{ab}=\kappa n^{ab}$, where $\kappa$ is the
surface gravity and $n^{ab}$ is the binormal to the horizon with the standard
normalization $n^{ab}n_{ab}=-2$)

\begin{equation}
  \label{eq:Iyer-Wald}
  \kappa \frac{\delta \mathbf{L}}{\delta R^{ab}}n^{ab}
  =
  \mathbf{Q}[k]+\ell_{s}P_{k}+\Delta_{a}k^{a}\,.
\end{equation}

\noindent
The second term in the right-hand side gives rise to a work term in the
thermodynamical relations and does not contribute to the entropy. The third
term vanishes when integrated over the bifurcation surface or when integrated
at spatial infinity, for stationary, asymptotically-flat black holes. Thus, we
expect the black-hole entropy to be given by the Iyer--Wald formula. As we
advanced, that the Iyer--Wald formula coincides the Lorentz Noether charge
given in Eq.~(\ref{eq:localLorentzcharge}) for $\sigma^{ab}= \kappa n^{ab}$,
and differs from the Noether--Wald charge in general.

\section{Scalar charges}
\label{sec-scalarcharges}

As we mentioned in Section~\ref{sec-global}, the charges obtained by
integrating the Noether 3-form currents $J_{m}$ and $J_{n}$ over closed
spacelike hypersurfaces in stationary black-hole space times, vanish
identically. However, although scalar charges are not conserved charges, they
are very useful to characterize the scalar field of black holes and occur in
the first law of black hole mechanics \cite{Gibbons:1996af}.  A manifestly
covariant definition of scalar charge that satisfies a Gauss law and which
reproduces the values of the conventional coordinate-dependent definition was
proposed in Refs.~\cite{Pacilio:2018gom,Ballesteros:2023iqb}. The fact that it
satisfies a Gauss law can sometimes be used to prove no-hair theorems.

The definition of these scalar charges is based on one fact and one
assumption: the on-shell closedness of the 3-form currents $J_{m}$ and $J_{n}$
and their invariance under the transformations generated by the Killing vector
$k$, $\delta_{k}$. For $J_{m}$, we find

\begin{equation}
  \begin{aligned}
    0
    & =
      \delta_{k}J_{m}
      =
      (-\pounds_{k}+\delta_{\Lambda_{k}})J_{m}
      \doteq
      -d\imath_{k}J_{m} -\tfrac{1}{2}\ell_{s}\delta_{\Lambda_{k}}C
    \\
    & \\
    & =
      d\left[-\imath_{k}\left(\star d\phi +e^{2\phi}\star d\chi\right)
      +\tfrac{1}{2}\ell_{s}P_{k} \right]
    \\
    & \\
    & \equiv
      d\mathbf{Q}_{m}[k]\,,
  \end{aligned}
\end{equation}

\noindent
and, for $J_{n}$,

\begin{equation}
  \begin{aligned}
    0
    & =
      \delta_{k}J_{n}
      =
      (-\pounds_{k}+\delta_{\sigma_{k}})J_{n}
      \doteq
      -d\imath_{k}J_{n} -\tfrac{1}{4}\delta_{\sigma_{k}} \omega_{(3)}^{L}
    \\
    & \\
    & =
      d\left[-\imath_{k}\left(\ell_{s}^{-2}e^{2\phi}\star d\chi - \tfrac{1}{4}\omega_{(3)}^{L}\right)
      -\tfrac{1}{4}\sigma_{k}{}^{a}{}_{b}d\omega^{b}{}_{a}\right]
    \\
    & \\
    & =
      d\left(-\ell_{s}^{-2}e^{2\phi}\imath_{k}\star d\chi
      -\tfrac{1}{2}P_{k\, ab}R^{ab}
      \right)
    \\
    & \\
    & \equiv
      d\mathbf{Q}_{n}[k]\,.
  \end{aligned}
\end{equation}

It is convenient to use \footnote{We are picking the linear combination of the charges $\mathbf{Q}_{n,m}$ which is diagonal with respect to the contribution due to the kinetic terms of the scalars.}

\begin{subequations}
  \begin{align}
    \mathbf{Q}_{\chi}[k]
    & \equiv
      \mathbf{Q}_{n}[k]\,,
    \\
    & \nonumber \\
  \mathbf{Q}_{\phi}[k]
  & \equiv
  \mathbf{Q}_{m}[k]-\ell_{s}^{2}\mathbf{Q}_{n}[k]
  =
  -\imath_{k}\star d\phi +\tfrac{1}{2}\ell_{s}P_{k}
      +\tfrac{1}{2}\ell_{s}^{2}P_{k\, ab}R^{ab}\,,
  \end{align}
\end{subequations}

\noindent
and, for the sake of simplicity, static, spherically-symmetric,
asymptotically-flat, black hole with metrics of the form

\begin{equation}
  ds^{2}
  =
  \lambda(r) dt^{2}-\lambda^{-1}(r)dr^{2}-R^{2}(r)d\Omega^{2}_{(2)}\,.
\end{equation}

The conventional definition of the dilaton ($\Sigma$) and axion ($\Upsilon$)
charges is given in terms of the asymptotic expansion of the fields by

\begin{subequations}
  \begin{align}
  \phi
  & \sim
  \phi_{\infty}+\Sigma/r+\mathcal{O}(r^{-2})\,,
    \\
    & \nonumber \\
  \chi
  & \sim
  \ell_{s}^{2}\left[\chi_{\infty}+e^{-2\phi_{\infty}}\Upsilon/r+\mathcal{O}(r^{-2})\right]\,.
  \end{align}
\end{subequations}

\noindent
and it is easy to see that, if $k=\partial_{t}$, the first terms in the
definitions of $\mathbf{Q}_{\phi}[k]$ and $\mathbf{Q}_{\chi}[k]$ give $\Sigma$
and $\Upsilon$ upon integration at spatial infinity, up to normalization:

\begin{equation}
  \begin{aligned}
  -\imath_{k}\star d\phi
  & =
  -k^{\mu}\frac{\varepsilon_{\mu\nu\rho\sigma}}{2\sqrt{|g|}}
    g^{\sigma\eta}\partial_{\eta}\phi dx^{\nu}\wedge dx^{\rho}
    \\
    & \\
    & =
  -\frac{\varepsilon_{t\theta\varphi r}}{R^{2}\sin{\theta}}
      g^{rr}\partial_{r}\phi d\theta\wedge d\varphi
    \\
    & \\
    & =
  -R^{2}\lambda \partial_{r}\phi \sin{\theta} d\theta\wedge d\varphi
    \\
    & \\
    & \sim
  \Sigma \omega_{(2)}\,,
  \end{aligned}
\end{equation}

\noindent
where $\omega_{(2)}$ is the volume form of the 2-sphere of unit radius
$S^{2}_{\infty}$, so

\begin{subequations}
  \begin{align}
    \Sigma
    & =
      -\frac{1}{4\pi}\int_{S^{2}_{\infty}} \imath_{k}\star d\phi\,,
    \\
    & \nonumber \\
    \Upsilon
    & =
      -\frac{1}{4\pi}\int_{S^{2}_{\infty}}\ell_{s}^{-2}
      e^{2\phi}\imath_{k}\star d\chi\,,
  \end{align}
\end{subequations}

The integrands we have used here differ from the 2-form charges
$\mathbf{Q}_{\phi}[k]$ and $\mathbf{Q}_{\chi}[k]$ by terms that vanish at
infinity,\footnote{$P_{k}$ is defined up to an exact 2-form that can be chosen
  so as to make $P_{k}$ vanish at spatial infinity.} and, thus, we can write

\begin{subequations}
  \begin{align}
    \Sigma
    & =
      \frac{1}{4\pi}\int_{S^{2}_{\infty}}\mathbf{Q}_{\phi}[k] \,,
    \\
    & \nonumber \\
    \Upsilon
    & =
  \frac{1}{4\pi}\int_{S^{2}_{\infty}}\mathbf{Q}_{\chi}[k]\,.
  \end{align}
\end{subequations}

The 2-form charges $\mathbf{Q}_{\phi}[k]$ and $\mathbf{Q}_{\chi}[k]$ are
closed on-shell and, therefore, the above integrals give the same value when
they are performed over smooth deformations of $S^{2}_{\infty}$, and, in
particular, over the bifurcation sphere, $\mathcal{BH}$. There, $k=0$, and,
using the well-known result

\begin{equation}
  \label{eq:Pkabproperty}
  P_{k}{}^{ab}
  \stackrel{\mathcal{BH}}{=}
  \kappa n^{ab}\,,
  \hspace{1cm}
  n^{ab}n_{ab}
  =
  -2\,,
\end{equation}

\noindent
where $n^{ab}$ is the binormal to the horizon,

\begin{subequations}
  \begin{align}
    \Upsilon
    & =
  -\frac{\kappa}{8\pi}\int_{\mathcal{BH}}R^{ab}n_{ab}\,,
    \\
    & \nonumber \\
    \Sigma
    & =
      \frac{\ell_{s}}{8\pi}\int_{\mathcal{BH}}
      P_{k}
      +\frac{\ell_{s}^{2}\kappa }{8\pi}\int_{\mathcal{BH}}
      R^{ab} n_{ab}
      =
      \frac{\ell_{s}}{8\pi}\int_{\mathcal{BH}}
      P_{k}
      -\ell_{s}^{2}\Upsilon\,.
  \end{align}
\end{subequations}

The first of these equations relates the axion charge to a geometric quantity
which has the structure of the gravitational charges constructed in
Ref.~\cite{Gomez-Fayren:2023qly} (times the surface gravity) as a
generalization of those constructed in Ref.~\cite{Benedetti:2023ipt}. Some of
these charges have been interpreted physically in
Refs.~\cite{Hull:2024xgo,Hull:2024mfb} as related to the dual graviton. It is
not one of them, though, because, in principle, $n_{ab}$ is only covariantly
constant over the bifurcation sphere and it is only defined over the Killing
horizon. If there was a covariantly constant bivector defined over the whole
black-hole spacetime whose restriction to the Killing horizon coincided with
the binormal, then we could apply Stokes theorem and show that this
gravitational charge satisfies a Gauss law. In any case, we can read the first
equation as a no-(primary)-hair \cite{Coleman:1991ku} theorem which restricts
the value of the scalar charge relating it to other quantities in the
solution.

The second equation relates the dilaton charge to the axion charge (hence, to
a geometrical quantity) and to the integral of the momentum map $P_{k}$, whose
interpretation is that of the chemical potential dual to $\ell_{s}$,
$\Phi_{\ell_{s}}$ \cite{Meessen:2022hcg}. Thus, we have

\begin{equation}
  \label{eq:Sigma}
  \Sigma
  =
  2\ell_{s}\Phi_{\ell_{s}}-\ell_{s}^{2}\Upsilon\,,
\end{equation}

\noindent
which can also be read as another no-(primary)-hair theorem for the dilaton.\footnote{The normalization of $\Phi_{\ell_{s}}$ is fixed requiring that $\ell_s$ is the associated charge and that the Smarr formula has the standard form.}

\section{Generalized Komar charge and Smarr formula}
\label{sec-Komar}

Having defined the conserved charges associated to the symmetries and the
scalar charges of the CR2 action we are ready to derive a Smarr formula for the
black-hole solutions of this theory using the techniques developed in
Refs.~\cite{Kastor:2008xb,Kastor:2010gq}.

We must first construct the generalized Komar charge associated to the Killing
vector $k$, $\mathbf{K}[k]$, which, in this case is simply given by
\cite{Komar:1958wp,Liberati:2015xcp,Hajian:2021hje,Ortin:2021ade,Mitsios:2021zrn,Meessen:2022hcg},

\begin{equation}
  \mathbf{K}[k]
  =
  -\mathbf{Q}[k]
  +\omega_{k}\,,
\end{equation}

\noindent
where $\omega_{k}$ is defined by

\begin{equation}
  \imath_{k}\left.\mathbf{L}\right|_{\rm on-shell}
  \equiv
  d\omega_{k}\,.
\end{equation}

In order to evaluate the Lagrangian on-shell, we take the trace of the Einstein
equation (\ref{eq:Ea}), which, in differential-form language, is

\begin{equation}
  \begin{aligned}
    e^{a}\wedge \mathbf{E}_{a}
    & =
      e^{a}\wedge \mathbf{E}^{(0)}{}_{a}
      +d\left(e^{a}\wedge \Delta_{a}\right)
    \\
    & \\
    & =
      -2\left[\mathbf{L}
      +\tfrac{1}{4}\ell_{s}^{2}\left(e^{-\phi} \tilde{R}^{ab}\wedge R_{ab}
      +\chi R^{ab}\wedge R_{ab}\right)
      -\ell_{s}dC
      \right]
      +d\left(e^{a}\wedge \Delta_{a}\right)
    \\
    & \\
    & =
      -2\left\{\mathbf{L}
+\mathbf{E}_{\phi} +d\star d\phi -e^{2\phi}d\chi \wedge \star d\chi
-\chi\left[\mathbf{E}_{\chi} +d\left(e^{2\phi}\star d\chi\right)\right]
      -\ell_{s}dC
      \right\}
    \\
    & \\
    & \hspace{.5cm}
      +d\left(e^{a}\wedge \Delta_{a}\right)\,,
  \end{aligned}
\end{equation}

\noindent
and, therefore

\begin{equation}
\left.\mathbf{L}\right|_{\rm on-shell}
  =
    \tfrac{1}{2}d\left(e^{a}\wedge \Delta_{a}\right)
  -d\star d\phi +e^{2\phi}d\chi \wedge \star d\chi
 +\chi d\left(e^{2\phi}\star d\chi\right)
 +\ell_{s}dC\,.
\end{equation}

Then,

\begin{equation}
  \begin{aligned}
\imath_{k}\left.\mathbf{L}\right|_{\rm on-shell}
  & =
    \tfrac{1}{2}\imath_{k}d\left(e^{a}\wedge \Delta_{a}\right)
  -\imath_{k}d\star d\phi +e^{2\phi}d\chi \wedge \imath_{k}\star d\chi
 +\chi \imath_{k}d\left(e^{2\phi}\star d\chi\right)
    +\ell_{s}\imath_{k}dC
    \\
    & \\
    & =
    -\tfrac{1}{2}d\imath_{k}\left(e^{a}\wedge \Delta_{a}\right)
  +d \imath_{k}\star d\phi -e^{2\phi}d\chi \wedge \imath_{k}\star d\chi
 -\chi d \left(e^{2\phi}\imath_{k}\star d\chi\right)
    -\ell_{s}dP_{k}
    \\
    & \\
    & =
d\left\{    -\tfrac{1}{2}\imath_{k}\left(e^{a}\wedge \Delta_{a}\right)
  +\imath_{k}\star d\phi -\chi e^{2\phi}\imath_{k}\star d\chi
    -\ell_{s}P_{k}\right\}\,,
  \end{aligned}
\end{equation}

\noindent
which gives $\omega_{k}$ and, then, taking into account our previous result for
the Noether--Wald charge Eq.~(\ref{eq:NWcharge}), the generalized Komar 2-form
charge is found to be given by

\begin{equation}
  \begin{aligned}
  \mathbf{K}[k]
  & =
  -\star (e^{a} \wedge e^{b})P_{k\, ab}
  -\tfrac{1}{2}\ell_{s}^{2}\left(e^{-\phi} \tilde{R}^{ab}
    +\chi R^{ab}\right)P_{k\, ab}
    +\tfrac{1}{2}\left(\Delta_{a}k^{a}+e^{a}\wedge \imath_{k}\Delta_{a}\right)
    \\
    & \\
    & \hspace{.5cm}
  +\imath_{k}\star d\phi -\chi e^{2\phi}\imath_{k}\star d\chi\,.
  \end{aligned}
\end{equation}

Using the equation of motion and adding total derivatives, this expression can
be substantially modified. we will stick with this one because the
interpretation of its integrals in terms of the black-hole charges is simpler.

\subsection{Smarr formula}

By construction, the generalized Komar charge is closed on-shell. Integrating
the equation $d\mathbf{K}[k]\doteq 0$ over a spacelike hypersurface
$\Sigma^{3}$ such that

\begin{equation}
  \partial\Sigma^{3}
  =
  \mathcal{BH}\cup \mathrm{S}^{2}_{\infty}\,,
\end{equation}

\noindent
and applying Stokes theorem taking into account the orientations, we find

\begin{equation}
  \int_{\mathcal{BH}}\mathbf{K}[k]
  =
  \int_{\mathrm{S}^{2}_{\infty}}\mathbf{K}[k]\,.
\end{equation}

Recovering the overall factor $(16\pi G_{N})^{-1}$, the integral at spatial
infinity gives

\begin{equation}
  \int_{\mathrm{S}^{2}_{\infty}}\mathbf{K}[k]
  =
  \tfrac{1}{2}M-\tfrac{1}{4}\left(\Sigma -\ell_{s}^{2}\chi_{\infty}\Upsilon\right)\,,
\end{equation}

\noindent
while, using Eq.~(\ref{eq:Pkabproperty}), the integral over the bifurcation
surface gives

\begin{equation}
  \int_{\mathcal{BH}}\mathbf{K}[k]
  =
  TS\,,
\end{equation}

\noindent
where

\begin{subequations}
  \begin{align}
    \label{eq:Hawkingtemperature}
    T
    & =
      \frac{\kappa}{2\pi}\,,
    \\
    & \nonumber \\
    \label{eq:Waldentropy}
    S
    & =
      -\frac{1}{8G_{N}}\int_{\mathcal{BH}}
\left\{        \star (e^{a} \wedge e^{b})
  +\tfrac{1}{2}\ell_{s}^{2}\left(e^{-\phi} \tilde{R}^{ab}
    +\chi R^{ab}\right)\right\}n_{ab}\,,
  \end{align}
\end{subequations}

\noindent
are the Hawking temperature and Wald entropy, respectively,
and the Smarr formula can be written in the form

\begin{equation}
  M
  =
  2TS+\tfrac{1}{2}\left(\Sigma -\ell_{s}^{2}\chi_{\infty}\Upsilon\right)\,,  
\end{equation}

\noindent
or, using Eq.~(\ref{eq:Sigma}), in the more conventional form

\begin{equation}
  M
  =
  2TS+\ell_{s}\Phi_{\ell_{s}}
  -\tfrac{1}{2}(1+\chi_{\infty})\alpha'\Upsilon\,.
\end{equation}

\noindent
The last two terms are of order $\alpha'$ and correspond to the deviation
from the standard Smarr formula that we had foreseen in the introduction. The
second of them is still unconventional because of its dependence on the
asymptotic value of the axion field. However, that dependence is just a
consequence of the invariance of the CR action under constant shifts of the
axion and of the fact that the Wald entropy is not: that term compensates for
those changes, keeping the left-hand side of the Smarr relation (the mass)
invariant under that transformation. 

In order to test these results, it is convenient to study some examples.

\subsection{The Smarr formula for $\alpha'$-corrected Schwarzschild  and Kerr
  black holes}

The Schwarzschild and Kerr solutions can be seen as particular solutions of
the CR theory at zeroth order in $\alpha'$ with constant scalars
$\phi=\phi_{\infty}$, $\chi=\chi_{\infty}$. At first order in $\alpha'$ these
scalars get corrected because of the $\mathcal{O}(\alpha')$ source terms,
evaluated over the zeroth-order solution. The scalars only enter the Einstein
equation (\ref{eq:Einsteinequation}) through their derivatives, which are
proportional to $\alpha'$ and, therefore, they only contribute to
$\mathcal{O}(\alpha^{\prime\, 2})$ terms. Thus, there are no
$\mathcal{O}(\alpha')$ corrections to the metrics, as discussed in the
introduction. 

Let us consider the Schwarzschild solution first, for the sake of simplicity.
The Pontrjagin 4-form vanishes, but the Gauss--Bonnet 4-form does not:

\begin{equation}
  GB_{\rm Schwarzschild}
  =
  -\frac{24 M^{2}\sin{\theta}}{r^{4}}dt\wedge dr\wedge d\theta\wedge d\varphi\,.
\end{equation}

Thus, only the dilaton is sourced and gets $\alpha'$ corrections and the
$\alpha'$-corrected Schwarzschild solution is given by\footnote{the
  integration constants have been chosen so as to have a dilaton which tends
  to a constant at infinity.}

\begin{subequations}
  \begin{align}
    ds^{2}
    & =
      \lambda dt^{2} -\lambda^{-1}dr^{2} -r^{2}d\Omega_{(2)}^{2}
      +\mathcal{O}(\alpha^{\prime\, 2})\,,
      \hspace{1cm}
      \lambda
      = 1 -\frac{2m}{r}\,,
    \\
    & \nonumber \\
    \phi
    & =
      \phi_{\infty}
      -\frac{\alpha' e^{-\phi_{\infty}}}{4m} \left(\frac{1}{r}
      +\frac{m}{r^{2}} +\frac{4m^{2}}{3r^{3}}\right)
      +\mathcal{O}(\alpha^{\prime\, 2})\,,
      \\
    \\
    & \nonumber \\
    \chi
    & =
      \chi_{\infty}
      +\mathcal{O}(\alpha^{\prime\, 2})\,.
  \end{align}
\end{subequations}

From this solution we can read the mass $M$, Hawking temperature $T$ and
scalar charge $\Sigma$

\begin{subequations}
  \begin{align}
    M
    & =
      m +\mathcal{O}(\alpha^{\prime\, 2})\,,
    \\
    & \nonumber \\
    T
    & =
      \frac{1}{8\pi m}+\mathcal{O}(\alpha^{\prime\, 2})\,,
    \\
    & \nonumber \\
    \Sigma
    & =
      -\frac{e^{-\phi_{\infty}}\alpha'}{4m}+\mathcal{O}(\alpha^{\prime\, 2})\,.
  \end{align}
\end{subequations}

\noindent
The entropy, given by the Wald entropy Eq.~(\ref{eq:Waldentropy}), is given by

\begin{equation}
  S
  =
  4\pi \left[m^{2} +\frac{e^{-\phi_{\infty}}\alpha'}{8}\right]+\mathcal{O}(\alpha^{\prime\, 2})\,,
\end{equation}

\noindent
and, therefore

\begin{equation}
  2TS
  =
  m +\frac{e^{-\phi_{\infty}}\alpha'}{8m} +\mathcal{O}(\alpha^{\prime\, 2})
  =
  M -\tfrac{1}{2}\Sigma +\mathcal{O}(\alpha^{\prime\, 2})
  =
  M -\ell_{s}\Phi_{\ell_{s}} +\mathcal{O}(\alpha^{\prime\, 2})
\end{equation}

\noindent
where

\begin{equation}
  \Phi_{\ell_{s}}
  =
  \frac{\Sigma}{2\ell_{s}}
  =
  -\frac{e^{-\phi_{\infty}}\ell_{s}}{8M} \,.
\end{equation}

If we vary $M,\phi_{\infty}$ and $\ell_{s}$ in the entropy, we find

\begin{equation}
  \begin{aligned}
  \delta S
  & =
  8\pi M \delta M
  -\frac{e^{-\phi_{\infty}}\alpha'}{2}\delta\phi_{\infty}
  +\frac{\pi e^{-\phi_{\infty}}\ell_{s}}{M} \delta \ell_{s}
    +\mathcal{O}(\alpha^{\prime\, 2})
    \\
    & \\
    & =
  8\pi M \left[ \delta M
  -\frac{e^{-\phi_{\infty}}\alpha'}{16 M}\delta\phi_{\infty}
  +\frac{e^{-\phi_{\infty}}\ell_{s}}{8 M} \delta \ell_{s}\right] 
    +\mathcal{O}(\alpha^{\prime\, 2})
    \\
    & \\
    & =
  \frac{1}{T} \left[ \delta M
  +\tfrac{1}{4}\Sigma \delta\phi_{\infty}
  -\Phi_{\ell_{s}}\delta \ell_{s}\right] 
    +\mathcal{O}(\alpha^{\prime\, 2})\,,
  \end{aligned}
\end{equation}

\noindent
which is a first law consistent with the identification of $\ell_{s}$ as a
charge with $\Phi_{\ell_{s}}$ as associated chemical potential. The sign of
the term proportional to $\delta\phi_{\infty}$, predicted in
Ref.~\cite{Gibbons:1996af} (see also Ref.~\cite{Ballesteros:2023muf}) is
purely conventional and differs from the one assigned in those references.

Let us now consider the Kerr solution in Boyer--Lindquist coordinates, whose
Pontrjagin 4-form does not vanish. The expressions we find contain many terms
and we have found it convenient to study the slow rotation approximation with
$J/M = a\ll 1$. At this order, the GB and P 4-forms are given by

\begin{subequations}
  \begin{align}
    GB
    & =
      \left[  -\frac{24 M^{2}}{r^{4}}
      +\frac{480 a^{2}m^{2}\cos^{2}{\theta}}{r^{6}}
      +\mathcal{O}(a^{4})
      \right] \sin{\theta}\,
      dt\wedge dr\wedge d\theta\wedge d\varphi \,,
    \\
    & \nonumber \\
    P
    & =
      a\left[ \frac{72 m^{2}}{r^{5}} -\frac{1200 a^{2}\cos^{2}{\theta}}{r^{7}}
      +\mathcal{O}(a^{4})\right]
      \sin{\theta}\cos{\theta}\, dt\wedge dr\wedge d\theta\wedge d\varphi\,.
  \end{align}
\end{subequations}

\noindent
Notice that the P 4-form goes to zero at infinity faster than the GB one. This
will make the axion tend to a constant at infinity faster than the dilaton,
which will result in a vanishing axion charge for a non-trivial axion field. 

Then, to this order in $a$ and first order in $\alpha'$, the
$\alpha'$-corrected Kerr solution is given by \cite{Cano:2021rey}

\begin{subequations}
  \begin{align}
    ds^{2}
    & =
      \left(1 -\frac{2mr}{r^{2}+a^{2}\cos^{2}{\theta}}\right) dt^{2}
      +\frac{4mar\sin^{2}{\theta}}{r^{2}+a^{2}\cos^{2}{\theta}}dtd\varphi
      -\frac{r^{2}+a^{2}\cos^{2}{\theta}}{r^{2}-2mr +a^{2}}dr^{2}
      \nonumber \\
    & \nonumber \\
    & \hspace{.5cm}
      -(r^{2}+a^{2}\cos^{2}{\theta}) d\theta^{2}
      -\left(r^{2}+a^{2}
      +\frac{2ma^{2}r\sin^{2}{\theta}}{r^{2}+a^{2}\cos^{2}{\theta}}\right)\,
      \sin^{2}{\theta}d\varphi^{2}
      +\mathcal{O}(\alpha^{\prime\, 2})\,,
    \\
    & \nonumber \\
    \phi
    & =
      \phi_{\infty}
      -\frac{\alpha' e^{-\phi_{\infty}}}{4m} \left(\frac{1}{r}
      +\frac{m}{r^{2}} +\frac{4m^{2}}{3r^{3}}\right)
      +\frac{\alpha'a^{2} e^{-\phi_{\infty}}}{16 m^{3}}\left[\frac{1}{r}
      +\frac{m}{r^{2}}+\frac{4 m^{2}}{5 r^{3}} +\frac{2m^{3}}{5 r^{4}}
      \right.
\nonumber     \\
    & \nonumber \\
    & \hspace{.5cm}
      \left.
      +\cos{\theta}^2\left(\frac{28m^{2}}{5 r^{3}} +\frac{84 m^{3}}{ 5r^{4}}
      +\frac{192 m^{4}}{ 5r^{5}}\right)\right]
      +\mathcal{O}(\alpha^{\prime\, 2},a^{2})\,,
      \\
    & \nonumber \\
    \chi
    & =
      \chi_{\infty}
      -\frac{\alpha' a e^{-2\phi_{\infty}}\cos{\theta}}{4}
      \left\{\frac{5}{4mr^{2}}+\frac{5}{2r^{3}}+\frac{9m}{2r^{4}}
      -a^{2}\left[ \frac{1}{8m^{3}r^{2}}
      +\frac{1}{4m^{2}r^{3}}  +\frac{3}{10m r^{4}}
      + \frac{1}{5r^{5}} 
      \right.\right.
    \nonumber \\
    & \nonumber \\
    & \hspace{.5cm}
      \left.\left.
      +\cos^{2}{\theta}\left(  \frac{3}{2mr^{4}} +\frac{6}{r^{5}}
      +\frac{50 m}{3 r^{6}} \right)
      \right]\right\}
      +\mathcal{O}(\alpha^{\prime\, 2},a^{4})\,,
  \end{align}
\end{subequations}

\noindent
and from it we can read

\begin{subequations}
  \begin{align}
    M
    & =
      m +\mathcal{O}(\alpha^{\prime\, 2})\,,
    \\
    & \nonumber \\
    J
    & =
      m a +\mathcal{O}(\alpha^{\prime\, 2})\,,
    \\
    & \nonumber \\
    \Omega_{H}
    & =
      \frac{a}{r_{+}+a^{2}} +\mathcal{O}(\alpha^{\prime\, 2})
      =
      \frac{a}{4m^{2}}\left[1
      +\frac{a^{2}}{4m^{2}}+\mathcal{O}(\alpha^{\prime\, 2},a^{4})\right]\,,
    \\
    & \nonumber \\
    T
    & =
      \frac{\sqrt{m^{2}-a^{2}}}{2\pi (r_{+}^{2}+a^{2})}
      +\mathcal{O}(\alpha^{\prime\, 2})
      =
      \frac{1}{8\pi m}\left(1- \frac{a^{2}}{4m^{2}}
      +\mathcal{O}(\alpha^{\prime\, 2},a^{4})\right)\,,
    \\
    & \nonumber \\
    \Sigma
    & =
      -\frac{e^{-\phi_{\infty}}\alpha'}{4m}
      \left[1 - \frac{a^{2}}{4m^{2}}+\mathcal{O}(\alpha^{\prime\, 2},a^{4})\right]\,.
    \\
    & \nonumber \\
    \Upsilon
    & =
      \mathcal{O}(\alpha^{\prime\, 2})\,.
  \end{align}
\end{subequations}

\noindent
where

\begin{equation}
  r_{+}
  =
  m+\sqrt{m^{2}-a^{2}}
  =
  2m\left[1 -\frac{a^{2}}{4m^{2}}+ \mathcal{O}(a^{4})\right]\,,
\end{equation}

\noindent
is the value of $r$ at which the event horizon lies.

We find that the entropy is given by

\begin{equation}
  S
  =
  4\pi \left(m -\frac{a^{2}}{4} +\frac{e^{-\phi_{\infty}}\alpha'}{8}\right)
  +\mathcal{O}(\alpha^{\prime\, 2},a^{4})\,,
\end{equation}

\noindent
and

\begin{equation}
  \begin{aligned}
  2TS
  & =
  m -\frac{a^{2}}{4m} +\frac{e^{-\phi_{\infty}}\alpha'}{16m}
  \left(1 -\frac{a^{2}}{4m^{2}}\right)
    +\mathcal{O}(\alpha^{\prime\, 2},a^{4})
    \\
    & \\
  & =
  M-2\Omega_{H}J-\ell_{s}\Phi_{\ell_{s}}
  +\mathcal{O}(\alpha^{\prime\, 2},a^{4})\,,
  \end{aligned}
\end{equation}

\noindent
where, again

\begin{equation}
  \Phi_{\ell_{s}} = \frac{\Sigma}{2\ell_{s}}\,.      
\end{equation}

Finally, varying the entropy with respect to $M,J,\phi_{\infty}$ and
$\ell_{s}$ we find the first law to the order at which we have truncated our
calculations:

\begin{equation}
  \delta S
  =
  \frac{1}{T} \left[ \delta M
    -\Omega_{H}\delta J
  +\tfrac{1}{4}\Sigma \delta\phi_{\infty}
  -\Phi_{\ell_{s}}\delta \ell_{s}\right] 
    +\mathcal{O}(\alpha^{\prime\, 2},a^{4})\,.
\end{equation}

\section{Conclusions}
\label{sec-discussion}

In this paper we have studied the thermodynamics of the black-hole solutions of
the Cano--Ruip\'erez version of the heterotic string theory effective action,
extended as in Ref.~\cite{Meessen:2022hcg} so as to have a spacetime-variable
$\alpha'$. This extension is necessary to correctly account for all the terms
occurring in the Smarr formula and first law, in which $\alpha'$ (or
$\ell_{s}$) plays the role of a thermodynamical variable as predicted in
Ref.~\cite{Hajian:2021hje,Ortin:2021ade} and as proven in
\cite{Zatti:2023oiq}. The Smarr formula we have derived also contains a very
peculiar term proportional to the asymptotic value of the axion and which is
associated to the dependence of the Wald entropy on that parameter.

We have also shown how the definitions of scalar charges proposed in
Ref.~\cite{Ballesteros:2023iqb} generalizing the proposal in
Ref.~\cite{Pacilio:2018gom} also allow us to prove no-(primary)-hair theorems
in this theory. Since in this theory the scalars are sourced by the
Gauss--Bonnet and Pontrjagin 4-forms it is not too surprising that the values
of the scalar charges allowed by these theorems in regular black holes with
bifurcate horizons have a geometrical meaning and are, actually related to the
gravitational charges defined in
Ref.~\cite{Gomez-Fayren:2023qly}. Unfortunately, in the examples considered,
the axion charge vanishes. However, it is not difficult to see\footnote{Work in
  progress.} that in the Taub--NUT solution it does not and it is (obviously)
related to the NUT charge. Since the NUT charge can be seen as the magnetic
mass (a charge to which the dual graviton would couple) the existence of
relation between these results and the results of
Refs.~\cite{Hull:2024xgo,Hull:2024mfb} is evident.  However, as in other cases
we have studied, although the structure of the charge is the same, the
parameters that actually occur in the expressions we find do not satisfy the
same properties as those that enter in the definitions of the charges and
ensure their closure. In particular, the Lorentz parameter that appears in our
expressions (the binormal to the horizon $n^{ab}$) is only covariantly
constant over the bifurcation sphere and the closure of the 2-form
$R_{ab}n^{ab}$ requires $n^{ab}$ to be covariantly constant outside the
bifurcation sphere. Nevertheless, we think that, since these expressions occur
naturally in the Wald entropy (multiplied by the constant values of the
scalars over the horizon), they contain relevant physical and geometrical
information and should be investigated further. In this sense, we notice that
in higher-dimensional theories containing terms of higher order in curvatures,
further generalizations of the charges constructed in
Ref.~\cite{Gomez-Fayren:2023qly} such as, for instance,

\begin{equation}
  R^{a}{}_{c}\wedge R^{cb}n_{ab}\,,
  \hspace{1cm}
  R^{ab}\wedge R^{cd}\varepsilon_{abcdef}n^{ef}\,,
\end{equation}

\noindent
which are covariantly constant when $n^{ab}$ is, will appear as natural
building blocks of the Wald entropy (in 6 dimensions, for the above examples).

Finally, the extension of the Cano--Ruip\'erez action that we have constructed
allows for a piecewise-constant $\alpha'$, regions with different values of
$\alpha'$ being separated by domain walls. Customarily, $\alpha'$ is always
taken as a constant in string theory and this possibility has never been
considered before, but the existence of domain walls associated to this
parameter and which do not fit in the standard spectrum of extended
perturbative and non-perturbative extended objects is a tantalizing
possibility worth exploring. 

\section*{Acknowledgments}

The work of TO and MZ has been supported in part by the MCI, AEI, FEDER (UE)
grants PID2021-125700NB-C21 (``Gravity, Supergravity and Superstrings''
(GRASS)) and IFT Centro de Excelencia Severo Ochoa CEX2020-001007-S. The work
of MZ has been supported by the fellowship LCF/BQ/DI20/11780035 from ``La
Caixa'' Foundation (ID 100010434). TO wishes to thank M.M.~Fern\'andez for her
permanent support.

\appendix

\section{Conventions and some identities}
\label{sec-conventions}

Our Levi-Civita spin connection and Lorentz curvature 2-form are defined by

\begin{subequations}
    \begin{align}
      \mathcal{D}e^{a}
      & =
        de^{a}-\omega^{a}{}_{b}\wedge e^{b}
        =
        0\,,
      \\
      & \nonumber \\
      R_{ab}
      & =
        d \omega_{ab} - \omega_{ac} \wedge \omega^{c}{}_{b}\, .      
    \end{align}
  \end{subequations}

\noindent
where $R_{abc}{}^{d}$ is the Riemann tensor,
$R_{\mathrm{ic}\, ab} \equiv R_{acb}{}^{c}$ is the Ricci tensor and
$R_{\mathrm{ic}} \equiv R_{\mathrm{ic}\, a}{}^{a}$ is the Ricci scalar.

Notice that

\begin{equation}
    \imath_{a}\tilde{R}^{ab}
    =
    \tfrac{1}{2}\varepsilon^{abcd}R_{a\, \mu\, cd}dx^{\mu}
    =
    0\,, 
\end{equation}

\noindent
due to the Bianchi identity. On the other hand,

\begin{subequations}
    \begin{align}
    \imath_{a}R^{ab}
    & =
    R_{a\mu}{}^{ab}dx^{\mu}
    =
    R_{\mathrm{ic}\,\mu}{}^{b}dx^{\mu}
    \equiv
    R_{\mathrm{ic}}{}^{b}\,,
    \\
    & \nonumber \\
    \imath_{a}R_{\mathrm{ic}}{}^{a}
    & =
    R_{\mathrm{ic}\,a}{}^{a}
    =
    R_{\mathrm{ic}}\,.
    \end{align}
\end{subequations}

The Einstein tensor 1-form is

\begin{equation}
    G^{a}
    \equiv
    R_{\mathrm{ic}}{}^{a} -\tfrac{1}{2}R_{\mathrm{ic}}e^{a}\,.
\end{equation}

\section{The building blocks of the equations of motion}
\label{app-buildingblocks}

Ignoring the $(16\pi G_{N})^{-1}$ factor, the building blocks of the equations
of motion and of the presymplectic potential defined implicitly in
Eq.~(\ref{eq:variationSCR2-1}) are given by

\begin{subequations}
  \begin{align}
    \label{eq:E0aexplicit}
    \mathbf{E}^{(0)}{}_{a}
    & =
      \imath_{a}\star (e^{b}\wedge e^{c})\wedge R_{bc}
      +\tfrac{1}{2}\left(\imath_{a}d\phi \star d\phi
      +d\phi\wedge \imath_{a}\star d\phi\right)
      \nonumber \\
    & \nonumber \\
    & \hspace{.5cm}
      +\tfrac{1}{2}e^{2\phi}\left(\imath_{a}d\chi \star d\chi
      +d\chi\wedge \imath_{a}\star d\chi\right)\,,
    \\
    & \nonumber \\
    \label{eq:Etilde1ab}
    \tilde{\mathbf{E}}^{(1)\,ab}
    & =
      -\star(e^{a}\wedge e^{b})
      -\tfrac{1}{2}\ell_{s}^{2}\left(e^{-\phi} \tilde{R}^{ab}
    +\chi R^{ab}\right)\,,
    \\
    & \nonumber \\
    \mathbf{E}^{(0)}{}_{\phi}
    & =
      e^{2\phi} d\chi \wedge \star d\chi
      +\tfrac{1}{4} \ell_{s}^{2}e^{-\phi} \tilde{R}^{ab}\wedge R_{ab}\,,
    \\
    & \nonumber \\
    \tilde{\mathbf{E}}^{(1)}{}_{\phi}
        & =
      -\star d\phi\,,
    \\
    & \nonumber \\
    \mathbf{E}^{(0)}{}_{\chi}
    & =
     -\frac{1}{4}\ell_{s}^{2} R^{ab}\wedge R_{ab} \,,
    \\
    & \nonumber \\
    \tilde{\mathbf{E}}^{(1)}{}_{\chi}
        & =
      -e^{2\phi}\star d\chi\,,
    \\
    & \nonumber \\
    \mathbf{E}_{\rm \ell_{s}}
        & =
          dC
          -\tfrac{1}{2}\ell_{s}\left(e^{-\phi} \tilde{R}^{ab}\wedge R_{ab}
    +\chi R^{ab}\wedge R_{ab}\right)\,,
    \\
    & \nonumber \\
    \tilde{\mathbf{E}}^{(1)}{}_{\rm C}
    & =
      \ell_{s}\,.
  \end{align}
\end{subequations}

Furthermore, $\mathbf{E}^{(1)}{}_{ab}$ and $\Delta_{a}$, defined in
Eq.~(\ref{eq:E1abdef}) and Eq.~(\ref{eq:Deltac}) respectively, and the
covariant derivative of the latter are given by

\begin{subequations}
  \label{eq:E1abexplicit}
  \begin{align}
    \mathbf{E}^{(1)\, ab}
    & =
      \tfrac{1}{2}\left[d\left(\ell_{s}^{2}e^{-\phi}\right) \tilde{R}^{ab}
      +d\left(\ell_{s}^{2}\chi\right) R^{ab}\right]\,,
    \\
    & \nonumber \\
    \Delta^{c}
    & =
      \imath_{a}d (\ell_{s}^{2} \, e^{-\phi} ) \tilde{R}^{ac}
      +\imath_{a}d ( \ell_{s}^{2} \chi) R^{ac}
      -d(\ell_{s}^{2} \chi) \wedge R_{\mathrm{ic}}{}^{c}
      -\tfrac{1}{2}\imath_{a}d(\ell_{s}^{2} \chi) G^{a}\wedge e^{c} \,,
    \\
    & \nonumber \\
    \mathcal{D}\Delta^{c}
    & =
      \mathcal{D}\imath_{a}d (\ell_{s}^{2} e^{-\phi} )\wedge \tilde{R}^{ac}
      + \mathcal{D}\imath_{a}d (\ell_{s}^{2} \chi )\wedge R^{ac}
      +d (\ell_{s}^{2} \chi )\wedge \mathcal{D} R_{\mathrm{ic}}{}^{c}
      \nonumber \\
    & \nonumber \\
    & \hspace{.5cm}
      -\tfrac{1}{2}\mathcal{D}\imath_{a}d (\ell_{s}^{2} \chi )\wedge G^{a}
      \wedge e^{c}
      -\tfrac{1}{2}\imath_{a}d (\ell_{s}^{2} \chi )\mathcal{D}G^{a}
      \wedge e^{c} \,,
  \end{align}
\end{subequations}

\noindent
and the presymplectic potential in Eq.~(\ref{eq:presymplecticpotential}) has
the explicit expression

\begin{equation}
  \label{eq:presymplecticpotentialexplicit}
  \begin{aligned}
    \mathbf{\Theta}(\varphi,\delta\varphi)
    & =
      -\left[\star(e^{a}\wedge e^{b})
      +\tfrac{1}{2}\ell_{s}^{2}\left(e^{-\phi} \tilde{R}^{ab}
      +\chi R^{ab}\right)\right] \wedge \delta \omega^{ab}
      +\star d\phi \, \delta \phi
      +e^{2\phi}\star d\chi \, \delta \chi
    \\
    & \\
    & \hspace{.5cm}
      +\ell_{s}\delta C
      +\Delta_{a}\wedge \delta e^{a}\,.
  \end{aligned}
\end{equation}

\section{The equations of motion}
\label{app-eoms}

With all the elements of the previous section we find the following explicit
forms for the equations of motion

\begin{subequations}
  \begin{align}
    \label{eq:Einsteinequation}
    \mathbf{E}_{c}
    & =
      \imath_{c} \star (e^{a} \wedge e^{b}) \wedge R_{ab} +
      \tfrac{1}{2}\left(\imath_{c} d\phi\star d\phi
      +d\phi \wedge\imath_{c}\star d \phi  \right)
      \nonumber \\
    & \nonumber \\
    & \hspace{.5cm}
      +\tfrac{1}{2}e^{2\phi}\left(\imath_{c}d\chi\star d\chi
      +d\chi\wedge\imath_{c}\star d\chi\right)
    \\
    & \nonumber \\
    & \hspace{.5cm}
      + 
      \left[ \mathcal{D}\imath_{a} d (\ell_{s}^{2} \, e^{-\phi} )\wedge \tilde{R}^{ac}
      + \mathcal{D}\imath_{a} d (\ell_{s}^{2} \, \chi)\wedge R^{ac}
      +d (\ell_{s}^{2} \, \chi)\wedge \mathcal{D} R_{\mathrm{ic}}{}^{c}
      \right.
      \nonumber \\
    & \nonumber \\
    & \hspace{.5cm}
      \left.
      -\tfrac{1}{2}\mathcal{D}\imath_{a} d(\ell_{s}^{2} \, \chi)\wedge G^{a}\wedge e^{c}
      -\tfrac{1}{2}\imath_{a} d(\ell_{s}^{2} \, \chi) \mathcal{D}G^{a}\wedge
      e^{c}\right]\,,
    \\
    & \nonumber \\      
    \mathbf{E}_{\phi}
    & =
      -d\star d \phi + e^{2\phi} d\chi \wedge \star d \chi
      +\tfrac{1}{4}\ell_{s}^{2}e^{-\phi} \tilde{R}^{ab} \wedge R_{ab}\,,
    \\
    & \nonumber \\
    \mathbf{E}_{\chi}
    & =
      - d \left(e^{2\phi} \star d \chi\right)
      -\tfrac{1}{4}\ell_{s}^{2}R^{ab} \wedge R_{ab} \,,
    \\
    & \nonumber \\
    \label{eq:Els}
    \mathbf{E}_{\ell_{s}}
    & =
      dC -\tfrac{1}{2}\ell_{s}\left(e^{-\phi} \tilde{R}^{ab}\wedge R_{ab}
      +\chi R^{ab}\wedge R_{ab}\right)\,,
    \\
    & \nonumber \\
    \mathbf{E}_{C}
    & =
      -d\ell_{s}\,.
	\end{align}
\end{subequations}


\begin{thebibliography}{99}

\bibitem{Cano:2021rey}
P.~A.~Cano and A.~Ruip\'erez,
``String gravity in D=4,''
Phys. Rev. D \textbf{105} (2022) no.4, 044022
\doi{10.1103/PhysRevD.105.044022}
[\arxiv{2111.04750} [hep-th]].
  
\bibitem{Bergshoeff:1989de}
E.~A.~Bergshoeff and M.~de Roo,
``The Quartic Effective Action of the Heterotic String and Supersymmetry,''
Nucl. Phys. B \textbf{328} (1989), 439-468
\doi{10.1016/0550-3213(89)90336-2}
  
\bibitem{Ortin:2015hya}
T.~Ort\'{\i}n,
``Gravity and Strings'', 2nd edition, 
Cambridge University Press, 2015.

\bibitem{Gomez-Fayren:2023qly}
C.~G\'omez-Fayr\'en, P.~Meessen and T.~Ort\'{\i}n,
``Covariant generalized conserved charges of General Relativity,''
JHEP \textbf{09} (2023), 174
\doi{10.1007/JHEP09(2023)174}
[\arxiv{2307.04041} [gr-qc]].

\bibitem{Sopuerta:2009iy}
C.~F.~Sopuerta and N.~Yunes,
``Extreme and Intermediate-Mass Ratio Inspirals
in Dynamical Chern-Simons Modified Gravity,''
Phys. Rev. D \textbf{80} (2009), 064006
\doi{10.1103/PhysRevD.80.064006}
[\arxiv{0904.4501} [gr-qc]].

\bibitem{Gibbons:1994ff}
G.~W.~Gibbons and R.~E.~Kallosh,
``Topology, entropy and Witten index of dilaton black holes,''
Phys. Rev. D \textbf{51} (1995), 2839-2862
\doi{10.1103/PhysRevD.51.2839}
[\hepth{9407118} [hep-th]].

\bibitem{Lee:1990nz}
J.~Lee and R.~M.~Wald,
``Local symmetries and constraints,''
J. Math. Phys. \textbf{31} (1990), 725-743
\doi{10.1063/1.528801}

\bibitem{Wald:1993nt}
R.~M.~Wald,
``Black hole entropy is the Noether charge,''
Phys.\ Rev.\ D {\bf 48} (1993) no.8,  R3427.
\doi{10.1103/PhysRevD.48.R3427}
[\grqc{9307038}].

\bibitem{Iyer:1994ys}
V.~Iyer and R.~M.~Wald,
``Some properties of Noether charge
and a proposal for dynamical black hole entropy,''
Phys. Rev. D \textbf{50} (1994), 846-864
\doi{10.1103/PhysRevD.50.846}
[\grqc{9403028} [gr-qc]].

\bibitem{Kastor:2008xb}
D.~Kastor,
``Komar Integrals in Higher (and Lower) Derivative Gravity,''
Class. Quant. Grav. \textbf{25} (2008), 175007
\doi{10.1088/0264-9381/25/17/175007}
[\arxiv{0804.1832} [hep-th]].

\bibitem{Kastor:2010gq}
D.~Kastor, S.~Ray and J.~Traschen,
``Smarr Formula and an Extended First Law for Lovelock Gravity,''
Class. Quant. Grav. \textbf{27} (2010), 235014
\doi{10.1088/0264-9381/27/23/235014}
[\arxiv{1005.5053} [hep-th]].

\bibitem{Liberati:2015xcp}
S.~Liberati and C.~Pacilio,
``Smarr Formula for Lovelock Black Holes: a Lagrangian approach,''
Phys. Rev. D \textbf{93} (2016) no.8, 084044
\doi{10.1103/PhysRevD.93.084044}
[\arxiv{1511.05446} [gr-qc]].

\bibitem{Ortin:2021ade}
T.~Ort\'{\i}n,
``Komar integrals for theories of higher order
in the Riemann curvature and black-hole chemistry,''
JHEP \textbf{08} (2021), 023
\doi{10.1007/JHEP08(2021)023}
[\arxiv{2104.10717} [gr-qc]].

\bibitem{Mitsios:2021zrn}
D.~Mitsios, T.~Ort\'{\i}n and D.~Pere\~niguez,
``Komar integral and Smarr formula for axion-dilaton
black holes versus S duality,''
JHEP \textbf{08} (2021), 019
\doi{10.1007/JHEP08(2021)019}
[\arxiv{2106.07495} [hep-th]].

\bibitem{Hajian:2021hje}
K.~Hajian, H.~\"Oz\c{s}ahin and B.~Tekin,
``First law of black hole thermodynamics and
Smarr formula with a cosmological constant,''
Phys. Rev. D \textbf{104} (2021) no.4, 044024
\doi{10.1103/PhysRevD.104.044024}
[\arxiv{2103.10983} [gr-qc]].

\bibitem{Meessen:2022hcg}
P.~Meessen, D.~Mitsios and T.~Ort\'{\i}n,
``Black hole chemistry, the cosmological constant and the embedding tensor,''
JHEP \textbf{12} (2022), 155
\doi{10.1007/JHEP12(2022)155}
[\arxiv{2203.13588} [hep-th]].

\bibitem{Elgood:2020svt}
Z.~Elgood, P.~Meessen and T.~Ort\'{\i}n,
``The first law of black hole mechanics in
the Einstein-Maxwell theory revisited,''
JHEP \textbf{09} (2020), 026
\doi{10.1007/JHEP09(2020)026}
[\arxiv{2006.02792} [hep-th]].

\bibitem{Elgood:2020mdx}
Z.~Elgood, D.~Mitsios, T.~Ort\'{\i}n and D.~Pere\~n\'{\i}guez,
``The first law of heterotic stringy black hole mechanics
at zeroth order in $\alpha'$,''
JHEP \textbf{07} (2021), 007
\doi{10.1007/JHEP07(2021)007}
[\arxiv{2012.13323} [hep-th]].

\bibitem{Elgood:2020nls}
Z.~Elgood, T.~Ort\'{\i}n and D.~Pere\~niguez,
``The first law and Wald entropy formula of heterotic
stringy black holes at first order in $\alpha'$,''
JHEP \textbf{05} (2021), 110
\doi{10.1007/JHEP05(2021)110}
[\arxiv{2012.14892} [hep-th]].

\bibitem{Cano:2022tmn}
P.~A.~Cano, T.~Ort\'\i{}n, A.~Ruip\'erez and M.~Zatti,
``Non-extremal, \ensuremath{\alpha}'-corrected black holes in 5-dimensional heterotic superstring theory,''
JHEP \textbf{12} (2022), 150
doi:10.1007/JHEP12(2022)150
[arXiv:2210.01861 [hep-th]].


\bibitem{Zatti:2023oiq}
M.~Zatti,
``$\alpha'$ corrections to 4-dimensional non-extremal stringy black holes,''
JHEP \textbf{11} (2023), 185
\doi{10.1007/JHEP11(2023)185}
[\arxiv{2308.12879} [hep-th]].

\bibitem{Smarr:1972kt}
L.~Smarr,
``Mass formula for Kerr black holes,''
Phys. Rev. Lett. \textbf{30} (1973), 71-73
[erratum: Phys. Rev. Lett. \textbf{30} (1973), 521-521]
\doi{10.1103/PhysRevLett.30.71}

\bibitem{Komar:1958wp}
A.~Komar,
``Covariant conservation laws in general relativity,''
Phys. Rev. \textbf{113} (1959), 934-936
\doi{10.1103/PhysRev.113.934}

\bibitem{Bergshoeff:2001pv}
E.~Bergshoeff, R.~Kallosh, T.~Ort\'{\i}n, D.~Roest and A.~Van Proeyen,
``New formulations of D = 10 supersymmetry and D8 - O8 domain walls,''
Class. Quant. Grav. \textbf{18} (2001), 3359-3382
\doi{10.1088/0264-9381/18/17/303}
[\hepth{0103233} [hep-th]].

\bibitem{Ballesteros:2023iqb}
  R.~Ballesteros, C.~G\'omez-Fayr\'en,
T.~Ort\'{\i}n and M.~Zatti,
 ``On scalar charges and black hole thermodynamics,''
JHEP \textbf{05} (2023), 158
\doi{10.1007/JHEP05(2023)158}
[\arxiv{2302.11630} [hep-th]].

\bibitem{Jacobson:2015uqa}
T.~Jacobson and A.~Mohd,
``Black hole entropy and Lorentz-diffeomorphism Noether charge,''
Phys. Rev. D \textbf{92} (2015), 124010
\doi{10.1103/PhysRevD.92.124010}
[\arxiv{1507.01054} [gr-qc]].

\bibitem{Ortin:2002qb}
T.~Ort\'{\i}n,
``A note on Lie-Lorentz derivatives,''
Class. Quant. Grav. \textbf{19} (2002), L143-L150
\doi{10.1088/0264-9381/19/15/101}
[\hepth{0206159} [hep-th]].

\bibitem{Gibbons:1996af}
G.~W.~Gibbons, R.~Kallosh and B.~Kol,
``Moduli, scalar charges, and the first law of black hole thermodynamics,''
Phys. Rev. Lett. \textbf{77} (1996), 4992-4995
\doi{10.1103/PhysRevLett.77.4992}
[\hepth{9607108} [hep-th]].

\bibitem{Pacilio:2018gom}
C.~Pacilio,
``Scalar charge of black holes in Einstein-Maxwell-dilaton theory,''
Phys. Rev. D \textbf{98} (2018) no.6, 064055
\doi{10.1103/PhysRevD.98.064055}
[\arxiv{1806.10238} [gr-qc]].

\bibitem{Benedetti:2023ipt}
V.~Benedetti, P.~Bueno and J.~M.~Mag\'an,
``Generalized Symmetries for Generalized Gravitons,''
Phys. Rev. Lett. \textbf{131} (2023) no.11, 111603
\doi{10.1103/PhysRevLett.131.111603}
[\arxiv{2305.13361} [hep-th]].

\bibitem{Hull:2024xgo}
C.~Hull, M.~L.~Hutt and U.~Lindstr\"om,
``Charges and topology in linearised gravity,''
JHEP \textbf{07} (2024), 097
\doi{10.1007/JHEP07(2024)097}
[\arxiv{2401.17361} [hep-th]].

\bibitem{Hull:2024mfb}
C.~Hull, M.~L.~Hutt and U.~Lindstr\"om,
``Gauge-invariant magnetic charges in linearised gravity,''
Class. Quant. Grav. \textbf{41} (2024) no.19, 195012
\doi{10.1088/1361-6382/ad718a}
[\arxiv{2405.08876} [hep-th]].

\bibitem{Coleman:1991ku}
S.~R.~Coleman, J.~Preskill and F.~Wilczek,
``Quantum hair on black holes,''
Nucl. Phys. B \textbf{378} (1992), 175-246
\doi{10.1016/0550-3213(92)90008-Y}
[\hepth{9201059} [hep-th]].

\bibitem{Ballesteros:2023muf}
R.~Ballesteros and T.~Ort\'{\i}n,
``Hairy black holes, scalar charges and extended thermodynamics,''
Class. Quant. Grav. \textbf{41} (2024) no.5, 055007
\doi{10.1088/1361-6382/ad210a}
[\arxiv{2308.04994} [gr-qc]].






\end{thebibliography}
\end{document}